\documentclass[10pt,twocolumn]{IEEEtran_v15}
\usepackage{graphicx}

\def\microk{$\mu{\mbox{K}}$}
\def\mathrelfun#1#2{\lower3.6pt\vbox{\baselineskip0pt\lineskip.9pt
  \ialign{$\mathsurround=0pt#1\hfil##\hfil$\crcr#2\crcr\sim\crcr}}}

\makeatletter
\def\ifundefined{\@ifundefined}
\makeatother

\begin{document}


\title{A Cosmic Microwave Background Radiation Polarimeter
Using Superconducting Bearings}

\author{Shaul Hanany, Tomotake Matsumura, Brad Johnson, Terry Jones, John
R. Hull, and Ki B.\ Ma\thanks{S. Hanany, T. Matsumura, B. Johnson and T. Jones 
are with the School of Physics and Astronomy at the University of 
Minnesota/Twin Cities, USA ~ (e-mail: hanany@physics.umn.edu). \newline
J. R. Hull is with Argonne National laboratory. \newline
K. Ma is with the Texas Center for Superconductivity at the University 
of Houston }}
\ifundefined{IEEEtransversionmajor}{%

   \newlength{\IEEEilabelindent}
   \newlength{\IEEEilabelindentA}
   \newlength{\IEEEilabelindentB}
   \newlength{\IEEEelabelindent}
   \newlength{\IEEEdlabelindent}
   \newlength{\labelindent}
   \newlength{\IEEEiednormlabelsep}
   \newlength{\IEEEiedmathlabelsep}
   \newlength{\IEEEiedtopsep}

   \providecommand{\IEEElabelindentfactori}{1.0}
   \providecommand{\IEEElabelindentfactorii}{0.75}
   \providecommand{\IEEElabelindentfactoriii}{0.0}
   \providecommand{\IEEElabelindentfactoriv}{0.0}
   \providecommand{\IEEElabelindentfactorv}{0.0}
   \providecommand{\IEEElabelindentfactorvi}{0.0}
   \providecommand{\labelindentfactor}{1.0}
   
   \providecommand{\iedlistdecl}{\relax}
   \providecommand{\calcleftmargin}[1]{
                   \setlength{\leftmargin}{#1}
                   \addtolength{\leftmargin}{\labelwidth}
                   \addtolength{\leftmargin}{\labelsep}}
   \providecommand{\setlabelwidth}[1]{
                   \settowidth{\labelwidth}{#1}} 
   \providecommand{\usemathlabelsep}{\relax}
   \providecommand{\iedlabeljustifyl}{\relax}
   \providecommand{\iedlabeljustifyc}{\relax}
   \providecommand{\iedlabeljustifyr}{\relax}
 
   \newif\ifnocalcleftmargin
   \nocalcleftmarginfalse

   \newif\ifnolabelindentfactor
   \nolabelindentfactorfalse
   
   \newif\ifcenterfigcaptions
   \centerfigcaptionsfalse
   
   \let\OLDitemize\itemize
   \let\OLDenumerate\enumerate
   \let\OLDdescription\description
   
   \renewcommand{\itemize}[1][\relax]{\OLDitemize}
   \renewcommand{\enumerate}[1][\relax]{\OLDenumerate}
   \renewcommand{\description}[1][\relax]{\OLDdescription}

   \providecommand{\pubid}[1]{\relax}
   \providecommand{\pubidadjcol}{\relax}
   \providecommand{\specialpapernotice}[1]{\relax}
   \providecommand{\overrideIEEEmargins}{\relax}
   
   \let\CMPARstart\PARstart 
   
   \let\OLDappendix\appendix
   \renewcommand{\appendix}[1][\relax]{\OLDappendix}
   
   \newif\ifuseRomanappendices
   \useRomanappendicestrue
   
   \let\OLDbiography\biography
   \let\OLDendbiography\endbiography
   \renewcommand{\biography}[2][\relax]{\OLDbiography{#2}}
   \renewcommand{\endbiography}{\OLDendbiography}
   
   \markboth{A Test for IEEEtran.cls--- {\tiny \bfseries
   [Running Older Class]}}{Shell: A Test for IEEEtran.cls}}{
  
   \markboth{A Test for IEEEtran.cls--- {\tiny \bfseries
   [Running Enhanced Class
    V\IEEEtransversionmajor.\IEEEtransversionminor]}}%
   {Shell: A Test for IEEEtran.cls}}

%
%

\maketitle


\begin{abstract}
Measurements of the polarization of the cosmic microwave background
(CMB) radiation are expected to significantly increase our
understanding of the early universe.  We present a design for a CMB
polarimeter in which a cryogenically cooled half wave plate rotates by
means of a high-temperature superconducting (HTS) bearing. The design
is optimized for implementation in MAXIPOL, a balloon-borne CMB
polarimeter.  A prototype bearing, consisting of commercially
available ring-shaped permanent magnet and an array of YBCO bulk HTS
material, has been constructed. We measured the coefficient of
friction as a function of several parameters including temperature
between 15 and 80 K, rotation frequency between 0.3 and 3.5 Hz,
levitation distance between 6 and 10 mm, and ambient pressure between
$10^{-7}$ and 1 torr.  The low rotational drag of the HTS bearing
allows rotations for long periods of time with minimal input power and
negligible wear and tear thus making this technology suitable for a
future satellite mission.

\end{abstract}

\setcounter{table}{0}

\section{Introduction}

\PARstart{S}{uperconducting} bearings are often considered one of the
major uses of bulk high-temperature superconductor (HTS), and
considerable development effort of such bearings has occurred since
the discovery of HTS \cite{hull_2000}.  HTS bearings are characterized
by very low drag torque.  This, together with stiffness and damping
against perturbations from equilibrium have led to investigations of
their use in flywheel energy systems 
\cite{mulcahyetal_2001}-\cite{dayetal_2002},
a lunar telescope \cite{leeetal_1999}, sensitive gravimeters
\cite{hulletal_grav_1999}, and as a sensitive detector of gas pressure
\cite{chewetal_1995}.  Here we present a design for a cosmic microwave
background (CMB) radiation polarimeter that is based on levitating a
rotating half wave plate by means of an HTS bearing.

There is intense scientific and popular interest in 
uncovering the physical conditions in the universe at epochs as close
as possible to the big bang.  Tremendous progress in this endeavor
has been achieved already through detailed measurements of the cosmic
microwave background radiation \cite{partridge}.  The CMB
permeates the entire universe and has a black body spectrum with an
equivalent temperature of 2.73 K \cite{mather_etal90}. Information
about the {\em spatial anisotropy} in the intensity of the CMB, as measured
by a number of recent experiments 
\cite{debernardis_etal2000}-\cite{halverson_etal2002}
has revealed that the universe is topologically flat
\cite{lange_etal2001}-\cite{pryke_etal2002}
and has lent strong support to the idea that the universe has
undergone a period of exponential growth, an 'inflation', at about
$10^{-34}$ seconds after the bang \cite{linde_inflation}.  By
combining the data from CMB and other astrophysical measurements we
can now determine that only 5\% of the matter and energy density in
the universe is made of ordinary electrons, quarks, neutrinos and
photons, and that the rest is split into $\sim 2/3$ 'vacuum energy'
and $\sim 1/3$ a new form of 'dark matter'. Aside from their
existence, very little is known about either the vacuum energy or the
dark matter.  CMB measurements have also helped to constrain the age
of the universe to within about 10\% accuracy and to construct a
reliable scenario for its evolution.

The \textit{polarization} of the CMB radiation field encodes new and
valuable information about the early universe and an intense research
effort has begun to discover and characterize the polarization
signal. Detection of the polarization signal will enhance our
confidence in the prevailing cosmological model and will provide more
information about the properties of the constituents of the universe 
\cite{Zal97}. More importantly, the polarization signal seems to be the only 
direct probe of the inflationary epoch close to the big bang
\cite{KamKosSte97a}.  A detection of such a signal will provide
strong supporting evidence for the concept of inflation and will
illuminate the physical processes involved.  

The polarization anisotropy of the CMB is expected to have two
components, the 'scalar' and 'tensor' components
\cite{KamKosSte97a}. The scalar component should have a magnitude of
about a few \microk, about a factor of 10 smaller than the 50
\microk\ RMS temperature anisotropy.  The tensor component, which has
its origin in the inflationary epoch, is smaller by {\em at least} a
factor of 100 compared to the temperature anisotropy. CMB receivers
are just now reaching the sensitivity required to detect the scalar
component.
A high signal-to-noise ratio detection of the CMB polarization will
require a combination of high sensitivity and strong rejection of
systematic errors.  An experimental approach that satisfies both of
these requirements is to use bolometric detectors at the focal plane
of an optical system that modulates the incoming polarization
signal. Bolometers are the most sensitive detectors at the frequency
range where the CMB is most intense \cite{lange2002}. Temporal
modulation of the incoming signal and subsequent lock-in is a standard
and powerful technique to separate small signals from noise. It is
also an effective technique to reject systematic errors that appear at
a frequency different from the signal modulation frequency.

The best technique currently available for modulating the polarization
of the incident radiation is by turning a half wave plate (HWP).  When
a HWP is placed in the path of the incident light and turned at a
rotation frequency $f$, the polarization vector of the output radiation
rotates at a frequency of $4f$. The radiation is then passed through a
'polarization analyzer', for example a fixed polarizing grid, and is
then detected. The signal recorded by the detector is modulated at a
frequency of $4f$ and standard lock-in techniques can be used to
extract its magnitude. Although a rotating HWP has been used with many
experiments, over a broad range of wavelengths 
\cite{jones_etal88}-\cite{leach_etal91}
the challenge in adapting the technique to CMB polarimetry is coupling
the rotation mechanism to bolometric detectors which have high
sensitivity to microphonic noise. Standard mechanical rotation
mechanisms, such as gears and bearings, have stick-slip friction which
induce vibrations. The vibrations deposit mechanical energy in the
detectors and cause wires to vibrate thereby inducing excess noise.

We have developed a magnetic bearing for use with polarization experiments
that is based on magnetic levitation of a HWP above a high temperature
superconductive (HTS) material. The apparatus is designed to fit in a
balloon borne bolometric experiment MAXIPOL that will attempt to
detect the scalar component of the CMB polarization \cite{maxipol}.  
The mechanism can be easily adapted to any polarization experiment
and particularly for a future NASA satellite mission dedicated for 
detailed mapping of the tensor component of the CMB polarization. 
We describe the
hardware in Section~\ref{sec:hardware}, report on initial tests in
Section~\ref{sec:results} and summarize in Section~\ref{sec:summary}. 

\section{Description of Hardware} 
\label{sec:hardware}

\subsection{Overview and Requirements}

\begin{figure}[t]
\centering
\includegraphics[width=3in]{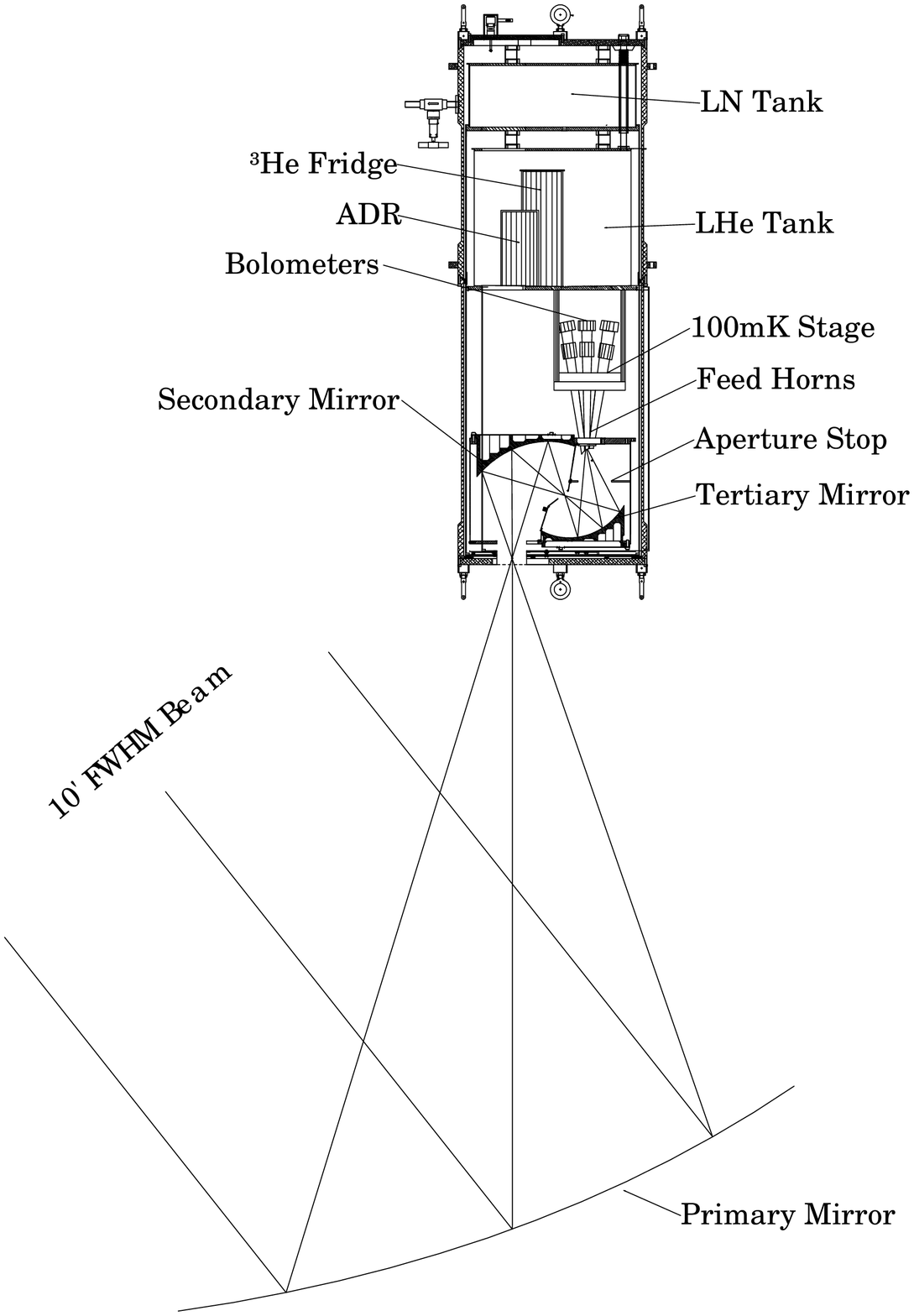}
\caption{
	Overall view of the MAXIPOL instrument. 
	}
\label{fig:maxima_system}
\end{figure}

MAXIPOL consists of an array of 16 bolometric photometers coupled to a
3 mirror off-axis Gregorian telescope, see
Fig.~\ref{fig:maxima_system}. The experiment is optimized for
stratospheric balloon borne ($\sim 38$ km) observations of the
CMB. The high altitude significantly reduces photon noise and noise
due to atmospheric fluctuations. The bolometers are maintained at a
temperature of 0.1 K by means of an adiabatic demagnetization
refrigerator \cite{hagmann}.  The optical system consists of
a 1.3 meter primary mirror and two additional mirrors that are
maintained at liquid helium temperature. 

Adapting a magnetic bearing system for a cryogenic balloon borne
experiment poses a number of requirements on the particular
implementation.
\begin{itemize}
\item The HWP must be cooled to liquid He temperatures to 
reduce the optical load on the detectors. 
\item The HWP must be located at the aperture stop of the optical 
system to eliminate a potential source of systematic error.
\item Characteristic rotation frequencies are $\sim 5$ Hz, limited by 
the $\sim 10$ msec response time of the bolometric detectors. 
\item It should be possible to accelerate the 
rotor periodically with minimal friction or electro-magnetic interference.  
\item The bearing system needs to be locked during 
periods of very strong accelerations, for example during launch or 
flight termination, and then released to begin operations. The latching
mechanism needs to be remote-controlled.  
\item The bearing system stiffness must be sufficiently high to 
operate through a range of tilt angles. Damping should be high to
dampen oscillations that may occur during typical observing
conditions.
\item All bearing properties must be stable over long periods 
of time. 

\end{itemize}

\subsection{The HWP and Magnetic Bearing Mechanism}

\begin{figure}[t]
\centering
\includegraphics[width=2.in]{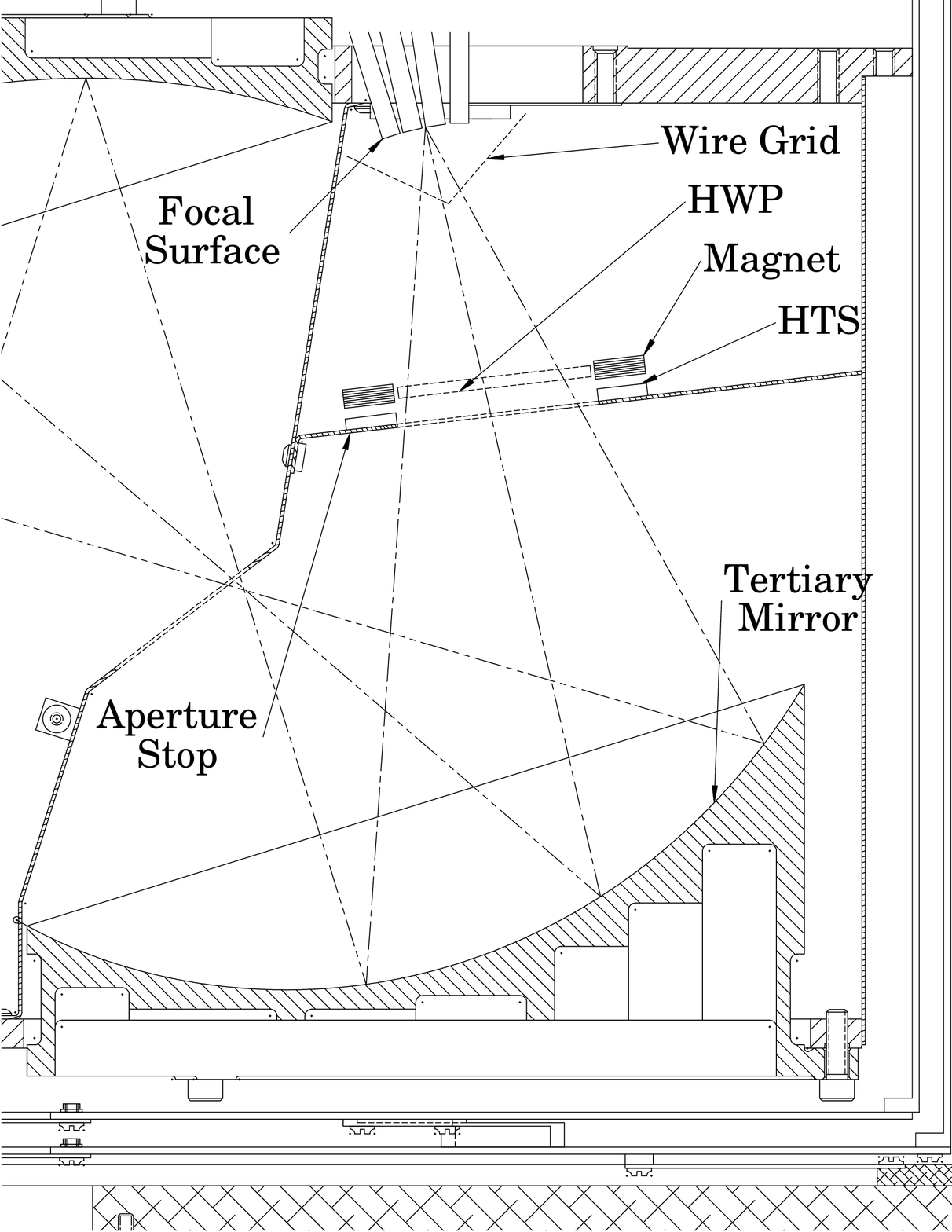}
\caption{
	The HWP and the magnetic bearing are located at the aperture 
	stop of the optical system, between the tertiary mirror and 
	the focal surface. A wire-grid polarizer, mounted at the 
	entrance aperture of the photometer array, is used to select  
	only one component of the incident polarization. 
	}
\label{fig:hts_hwp}
\end{figure}

The HWP and its rotation mechanism are located at the aperture stop of
the optical system, which is between the tertiary mirror and the focal
surface of the system, see Fig.~\ref{fig:hts_hwp}.  A
polarizing wire grid mounted at the entrance to the photometers
transmits only one component of the incident polarization vector.

\begin{figure}[t]
\centering
\includegraphics[width=2.5in,angle=-90]{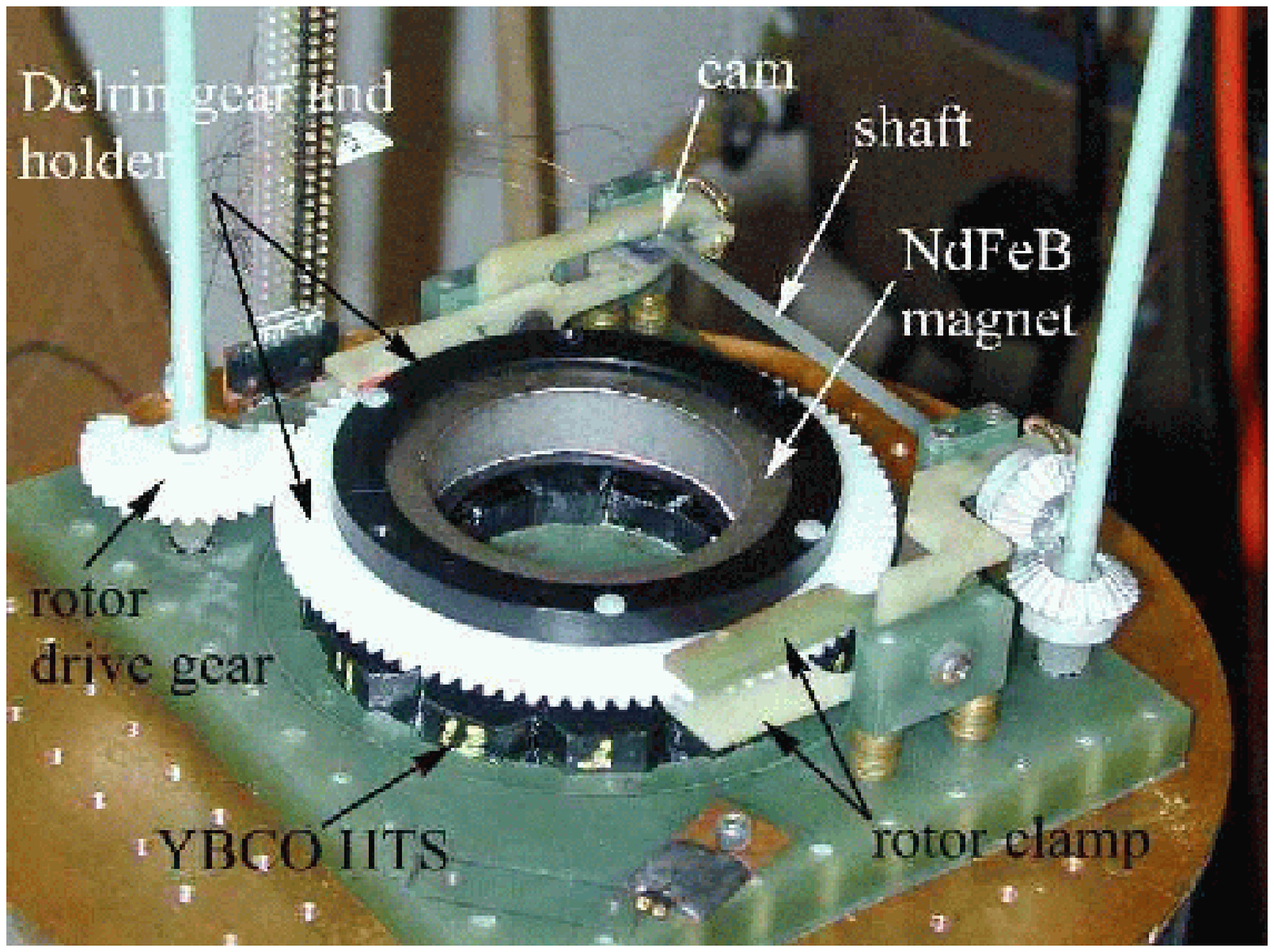}
\caption{
	The magnetic bearing apparatus during testing in a liquid 
	helium cryostat. The implementation in the MAXIPOL cryostat
	will be almost identical.  The rotor drive and
	rotor clamp gears will be driven with remote-controlled 
	stepping motors. 
	}
\label{fig:mbearing}
\end{figure}

The HWP consists of a 3.2 mm thick z-cut sapphire disk with a radius of
2.54 cm which is mounted inside a permanent ring-shaped magnet. The
magnet and the HWP are the rotor of a magnetic bearing that is
levitated above a ring of YBCO HTS material.  The sintered NdFeB magnet has an inside
radius of 2.54 cm, an outside radius of 3.56 cm, thickness of 1.2 cm,
and mass of 0.2 kg. It is polarized in the axial direction and 
has a remnance of $\sim 11\times 10^{3}$ Gauss and an energy
product of $30 \times 10^{6}$ Gauss-Oersted.  The moments of inertia
of the HWP and magnet are 83 and 1910
gr$\cdot$cm$^{2}$, respectively.  A HWP holder, made of Delrin and
with a gear at its outer circumference, holds the magnet/HWP
combination together and is part of the rotor, see
Fig.~\ref{fig:mbearing}.

The magnet is held at an appropriate distance $d$ above a ring of HTS
material which consists of 12 pieces of melt-textured YBCO
\cite{supercon}. The distance $d$ is a free parameter 
and is typically between 4 to 10 mm.  Two clamps, each
resembling a plier, hold the rotor in place during the
cool-down of the system.  A cryogenic stepping motor that drives a
shaft and a pair of cams is used to open and close the clamps.  The
HTS material and the clamps are mounted on a G-10 board.

Rotation of the rotor is achieved by means of a half gear that is
driven by a second stepping motor that is mounted outside of the
cryostat.  During cool-down the half gear and the gear at the outer
circumference of the rotor are engaged.  Once the system has cooled to
4 K, the clamps are opened and the half gear is turned with a
pre-programmed acceleration. 

If the need arises to re-rotate the rotor, the half-gear can be stepped 
slowly until it engages with the rotor, and the process repeats. 

The axial spring constant at 77 K was measured to be 256 and 136 N/m
for levitation distances of 8.6 and 10.7 mm, respectively. The
corresponding resonant frequencies are about 5.5 and 4 Hz,
respectively.

We tested all the mechanical components of the HTS bearing at liquid
nitrogen, liquid helium and intermediate temperatures. We carried out
spin-down measurements between $\sim 3.5$ and $\sim 0.3$ Hz to
determine the coefficient of friction of the bearing at various
temperatures, levitation heights, ambient pressures, and as a function
of the magnetic field structure of the magnet.  Measurements between
4.2 and 77 K were conducted in a liquid He cryostat (LHC) in which the
stator was mounted on a 0.9 cm thick copper cold-plate. The
temperature of the HTS was controlled with heating resistors and
measured with a calibrated resistance and diode thermometers.  The
rotation rate was measured with an LED and a photodiode.

\section{Tests, Results and Interpretation}
\label{sec:results}

Another set of spin-down measurements at 77 K were conducted in a
liquid nitrogen cryostat (LNC) with the apparatus described by Hull
{\em et al.} \cite{hulletal_1995}.  This apparatus was designed to
minimize energy losses due to eddy currents.  The measurements with
the LNC are summarized in Table~\ref{tab:meas_summary}.  In all
measurements the rotor consisted of the magnet without the HWP.

We quantify the results of the spin-down measurements in terms 
of the coefficient of friction (COF) \cite{hulletal_1995} given by 
\begin{equation}
\mbox{COF} = - C \alpha, 
\end{equation}
where $C = 2.7\times10^{-3}$ sec$^{2}$ is a rotor specific constant, 
and $\alpha$ is the angular acceleration. 

The angular acceleration $\alpha$ can be a function of the frequency 
of rotation $f$ and temperature $T$ and we parameterize it 
as (\cite{zeisberger_gawalek98}) 
\begin{equation}
\alpha (T) = 2\pi {df \over dt} = a_{0}(T) + 2 \pi a_{1}(T) f.  
\end{equation}
The coefficient $a_{0}$ has been interpreted as the contribution of
hysteresis to the angular acceleration and $a_{1}$ quantifies the
contributions from eddy currents and ambient pressure
\cite{hulletal_1995}. If $a_{1}$ is non-zero then $f$ should be an
exponential function of time.
\begin{table*}
\begin{center}
\caption{Summary of measurements with the LNC. In measurement 9 the magnetic field
of the magnet has been altered to reduce inhomogeneity by shimming with high permeability shims
(see text). }
\begin{tabular}{|c|c|c|c|c|c|} \hline
measurement & distance & Pressure & COF                   & $a_{0}$       & $a_{1}$ \\ 
   &  ($\pm 0.2$ mm) & (torr) &              & (rad/sec$^{2}$) & (1/sec)  \\ \hline
1   &  8.6     & $8\times 10^{-7}$ & $3.7\times 10^{-6}$ & $1.4\times 10^{-3}$ & 0 \\ \hline
2   &  8.6     & $3\times 10^{-4}$ &        & $1.4\times 10^{-3}$ & $1.2\times10^{-4}$ \\ \hline
3   &  8.6     & $3\times 10^{-3}$ &        & $1.4\times 10^{-3}$ & $2.2\times10^{-4}$ \\ \hline
4   &  9.5     & $7\times 10^{-7}$ & $1.8\times 10^{-6}$ & $6.6\times 10^{-4}$ & 0 \\ \hline
5   &  10.7   & $8\times 10^{-7}$   & $1.3\times 10^{-6}$ & $4.9\times 10^{-4}$ & 0 \\ \hline
6   &  10.7   & $3\times 10^{-4}$ &        & $4.9\times 10^{-4}$ & $7.9\times10^{-4}$ \\ \hline
7   &  10.7   & $8\times 10^{-4}$ &        & $4.9\times 10^{-4}$ & $8.9\times10^{-4}$ \\ \hline
8   &  10.7   & $3\times 10^{-3}$ &        & $4.9\times 10^{-4}$ & $1.1\times10^{-3}$ \\ \hline \hline
9   & 10.5   & $8\times 10^{-7}$ & $6.5\times 10^{-7}$ & $2.4\times 10^{-4}$ & 0  \\ \hline 
\end{tabular}
\end{center}
\label{tab:meas_summary}
\end{table*}

\subsection{COF Vs. Distance at 77 K: Hysteresis Losses}
\label{sec:cof_vs_dist}

A comparison of lines 1,4,5 in Table~\ref{tab:meas_summary} gives a
comparison of the COF as a function of distance from the HTS with low
ambient pressures, and negligible contribution from eddy currents.
  
\begin{figure}
\centering
\includegraphics[width=3.5in]{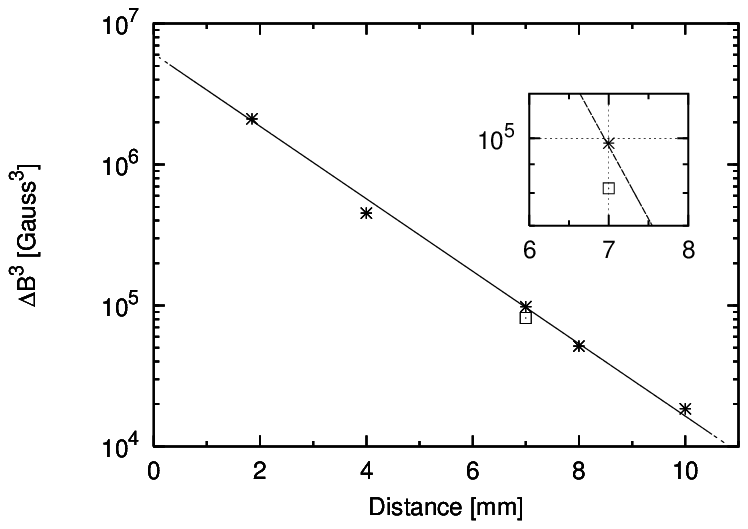}
\caption{
	The inhomogeneity of the magnetic field of the magnet as a
	function of distance from its surface for the bare magnet
	(stars) and for a configuration where the magnet is shimmed
	with high permeability steel (square).  
	The inhomogeneity is quantified using~(\ref{eqn:deltab3})
	and~(\ref{eqn:deltab}). The insert zooms
	on the difference of $1.65\times 10^{4}$ gauss$^{3}$ 
	between the shimmed and non-shimmed configurations.  
	The straight line is an approximate fit
	to the non-shimmed data and was used to find interpolated values.
	}
\label{fig:b3_vs_height}
\end{figure}

The COF decreases as the distance between the rotor and stator
increases.  This decrease is correlated well with a decrease in the
inhomogeneity of the axial component of the magnetic field as the
distance from the magnet increases.  We measured the strength of the
magnetic field as a function of distance from the surface of the
magnet. Measurements were taken at the inner, middle and outer radii
of the magnet and every 10 degrees in azimuth. Since in these
measurements the major contribution to the coefficient of friction is
expected to be hysteresis loss in the HTS, we follow Hull {\em et al}
\cite{hulletal_1996} and quantify the inhomogeneity of the field by
forming the quantity
\begin{equation}
\label{eqn:deltab3}
\Delta B^{3}(d) \equiv 
{ \sum_{i} \omega_{i} \left[\delta B_{z}^{3}(\theta,r_{i},d) \right]
\over \sum_{i} \omega_{i} }, 
\end{equation}
where $i$ runs over the inner, middle and outer radii and 
\begin{equation}
\label{eqn:deltab}
\delta B_{z}(\theta,r_{i}) =
\left|B_{z,\mbox{max}}(\theta,r_{i},d) - B_{z,\mbox{min}}(\theta,r_{i},d)\right|.
\end{equation}
The variables $\omega_{i}$ are weights that are proportional to the ratios of
elements of area as a function of radius. Figure~\ref{fig:b3_vs_height}
shows the quantity $\Delta B^{3}$ as a function of $d$. The reduction 
in magnetic field inhomogeneity with distance is evident. 

\begin{figure}
\centering
\includegraphics[width=3.5in]{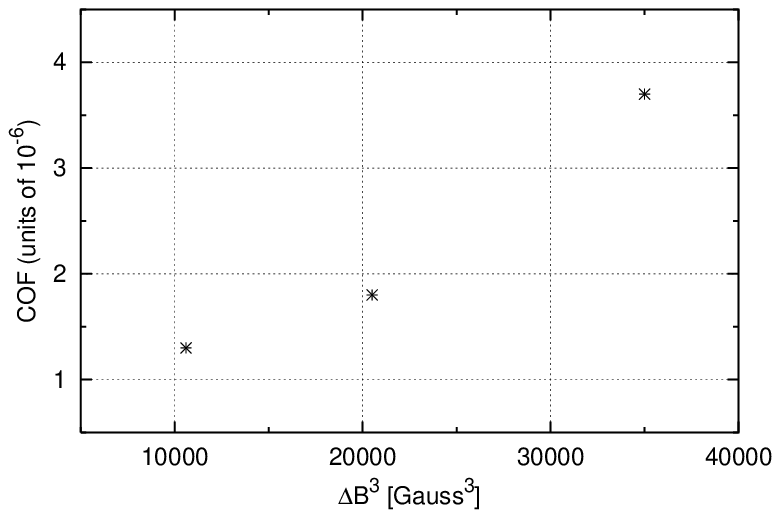}
\caption{
	The COF as a function of magnetic field inhomogeneity
        [from~(\ref{eqn:deltab3})].  The data suggests a linear
        relation as predicted by theory.
	}
\label{fig:cof_vs_b3}
\end{figure}

By interpolating the data of $\Delta B^{3}$ vs. height
(Fig.~\ref{fig:b3_vs_height}) we plot in Fig.~\ref{fig:cof_vs_b3} the COF
as a function of $\Delta B^{3}$. The number of data points is small,
nevertheless the data is suggestive of the linear relationship
predicted by theory \cite{hulletal_1996}.

In order to reduce hysteresis losses and decrease the coefficient of
friction we added shims in various locations on the surface of the
magnet that pointed toward the HTS \cite{hulletal_1996}. The shims
were made of 25 - 150 $\mu$m thick high permeability steel leaf.
Measurements of the magnetic field at a distance of 7 mm away from the
magnet gave less magnetic field inhomogeneity, as shown in
Fig.~\ref{fig:b3_vs_height}.  The COF was measured for a different
shimming configuration (for which measurements of $\Delta B^{3}$
are not available) and gave the lowest COF measured with the LNC, see
measurement 9.

\subsection{COF Vs. Pressure}
\label{sec:cof_vs_press}

\begin{figure}
\centering
\includegraphics[width=3.5in]{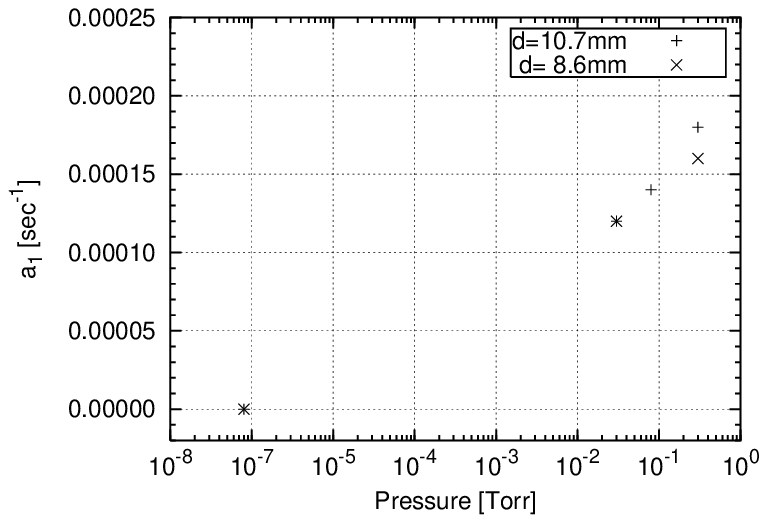}
\caption{
	The coefficient of friction as a function of ambient pressure
	for two levitation distances. All data was measured at 
	a temperature of 77 K. 
	}
\label{fig:a1_vs_press}
\end{figure}

Fig.~\ref{fig:a1_vs_press} shows the magnitude of the coefficient
$a_{1}$ as a function of ambient pressure. The measurement is a
compilation of data from two levitation distances. The data show
a decrease of $a_{1}$ with ambient pressure and are in broad agreement
with those of Weinberger et al. \cite{weinbergeretal_1991}, which 
suggest that $a_{1}$ becomes negligible at pressures below 
$\sim 5 \times 10^{-5}$ torr. 

\subsection{COF Vs. Temperature}
\label{sec:cof_vs_temp}

We used the LHC to measure the COF at levitation distances of 6 and
7.2 mm for various temperatures between 15 and 80 K and for
frequencies between $\sim 0.3$ and 3.5 Hz.  The magnet was released to
levitate when the helium cold plate was at $\sim 50$ K and we have no
information about the temperature of the magnet subsequently.

\begin{figure}
\centering
\includegraphics[width=3.5in]{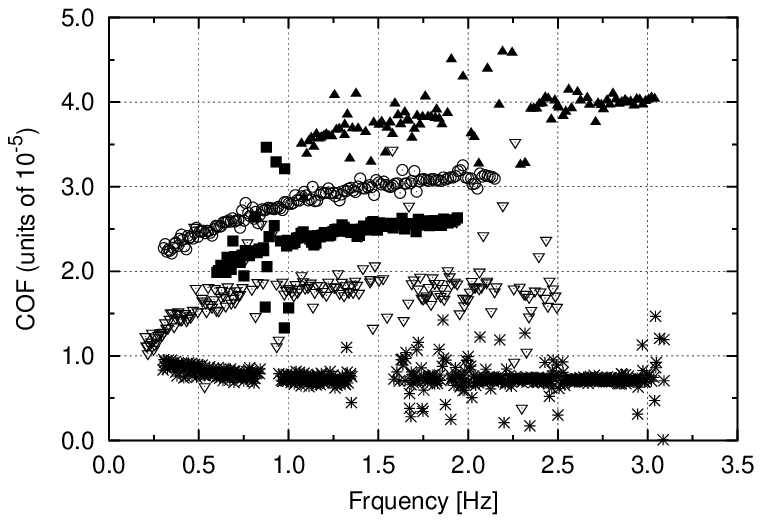}
\caption{
	The coefficient of friction as a function of frequency
	for temperatures of 16 K (stars), 60K (open triangles),
	70 K (squares), 79 K (circles), 84 K (filled triangles)
	for a levitation distance of 6.0 mm. For clarity of presentation, 
	the COF for temperatures
	of 70 K, 79 K, and 84 K, have been offset vertically up by
	0.4, 0.8, and 1.2 ordinate units, respectively. 	
	}
\label{fig:cof_vs_freq6mm}
\end{figure}

\begin{figure}
\centering
\includegraphics[width=3.5in]{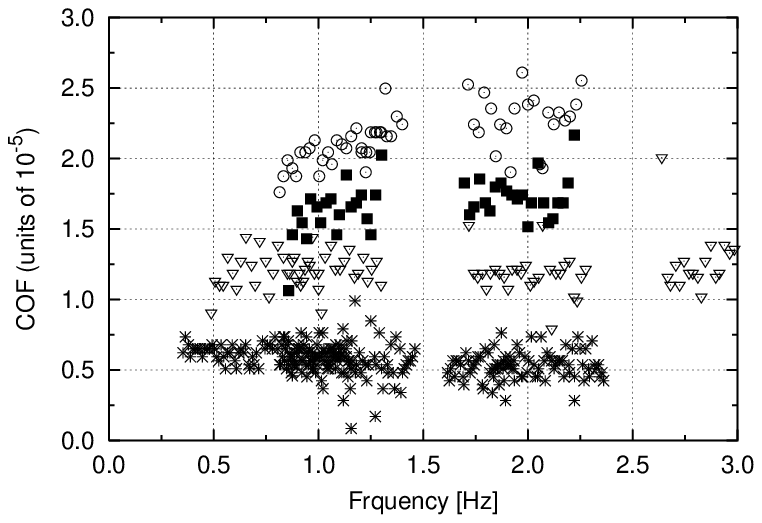}
\caption{
	The coefficient of friction as a function of frequency for 
	temperatures of 15 K (stars), 50 K (triangles), 62 K (squares)
	and 79 K (circles), for a levitation distance of 7.2 mm. For 
	clarity of presentation, the COF for temperatures of 62 and 
	79 K have been offset vertically up by 0.3 and 0.8 ordinate
	units, respectively. 
	}
\label{fig:cof_vs_freq7.2mm}
\end{figure}

Measurements of the COF as a function of rotation frequency are shown
in Figs~\ref{fig:cof_vs_freq6mm}
and~\ref{fig:cof_vs_freq7.2mm}.  At low temperatures ($\sim 15$ K)
the data were consistent with a constant deceleration for most
rotation frequencies with perhaps a slight increase in COF at
frequencies below 1 Hz.  At high temperatures, above about 60 K, the
data show an increasing COF with frequency.  The data at a
levitation distance of 6 mm and 60 K show a transition between a
constant COF above 1 Hz to a decreasing COF below 1 Hz.

\begin{figure}
\centering
\includegraphics[width=3.5in]{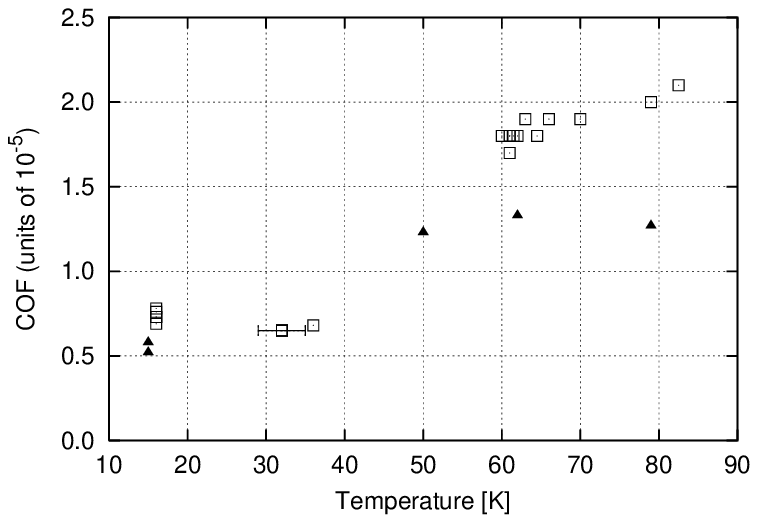}
\caption{
	The coefficient of friction at 1 Hz as a function of
	temperature for a levitation distance of 6 mm (squares) and 7.2
	mm (triangles). In one measurement, shown with a horizontal
	error bar, the temperature varied between 29 and 35 K.
	}
\label{fig:cof_vs_temp1hz}
\end{figure}

\begin{figure}
\centering
\includegraphics[width=3.5in]{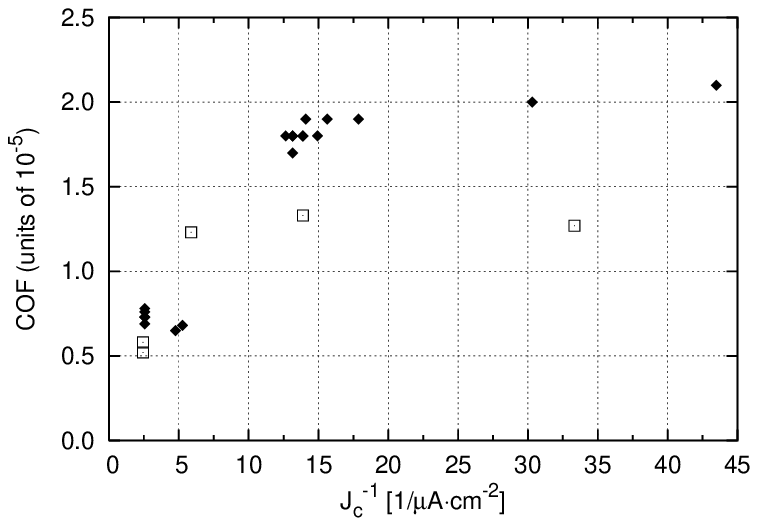}
\caption{
	The coefficient of friction at 1 Hz as a function of $1/J_{c}$, 
	where $J_{c}$ is the critical current in the HTS. 
	}
\label{fig:cof_vs_jc}
\end{figure}

Fig.~\ref{fig:cof_vs_temp1hz} shows the COF as a function of
temperature at a frequency of 1 Hz. A decrease in the COF by about a
factor of $\sim 3$ is evident for both levitation heights.
Hull et al. \cite{hulletal_1996} argue that the COF should be inversly
proportional to the critical current $J_{c}$ in the HTS.  Using data
from Zeisberger et al.\footnote{The data of Zeisberger et al.
\cite{zeisberger_gawalek98} is for 0, 1, and 2 Tesla and shows a weak
dependence on magnetic field strength. In our case the magnetic field
is $\sim 0.1$ T and we use the 0 T data.} \cite{zeisberger_gawalek98}
which give information about the critical current as a function of
temperature we plot the COF as a function of $1/J_{c}$ in
Fig.~\ref{fig:cof_vs_jc}. Since we don't have data about the
critical current as a function of temperature for {\em our} HTS
samples, it is premature to conclude that there is a disagreement
between the data and the model.

\section{Discussion and Summary}
\label{sec:summary}

We have presented a working prototype for an HTS bearing for use with
polarimeters that employ a HWP.  The complete absence of stick-slip
friction makes the magnetic bearing suitable for detector systems that
are sensitive to microphonic excitation. 

Our measurements indicate that with a readily available shimmed magnet
and HTS, at the range of gas pressures expected in a liquid He
cryostat and with no particular efforts to reduce eddy current losses
we can expect a COF of $\sim 3\times 10^{-6}$ (factor 2 less than the
measured COF at 15 K due to shimming), which would give rise to an
angular deceleration of about $1.7\times10^{-4}$ Hz/sec.  If the rotor
is set in motion and then slows down over a frequency range of 20 Hz,
suitable for CMB observations between 21 and 1 Hz given the time
constant of current bolometeric detectors, the bearing would need to
be reset in motion once in 33 hours. Since re-setting the bearing in
motion should take only a few minutes, the duty cycle of such a magnetic
bearing system is close to 100\%. The present design could be 
implemented in ground based or short duration balloon polarimeters. 

A COF of $\sim 3\times 10^{-6}$ is still two orders of magnitude
larger than the smallest COF reported for the 300-400 gr mass category
\cite{hull_2000}, indicating that a more careful design of the
bearing and the cryostat in which it operates can provide
uninterrupted rotation for more than 4 months.  Even just a factor of
10 improvement in the COF would give a continuous rotation for a
month, more than adequate for a long duration balloon mission.  

With a slowly decelerating rotor the frequency where the polarization
signal appears will also decrease. This could provide a powerful check
on systematic errors, but could also complicate the analysis of the
data.  Rather than let the bearing slow down over a long period of
time, it could be driven continuously with an induction or standard
brushless motor \cite{zhangetal_2002}.  If necessary a feed-back loop
can be implemented to maintain constant speed. Because of the small
COF, minimal power is required to drive such a motor and this approach
appears most attractive for a future satellite polarimetry mission.

Despite of the success of our prototype more testing needs to be done
to characterize the bearing performance.  In particular it is
essential to measure bearing oscillations, particularly at low
temperatures where hysteretic damping is lower compared to 77 K, and
the stability of the bearing parameters, e.g. stiffness, and
levitation distance, should be tested over a long period of time.


\begin{thebibliography}{11}

\bibitem{hull_2000}
J. Hull, "Superconducting bearings," {\it Supercond. Sci. Technol.}, 
Vol. 13, pp. R1 - R14, 2000.

\bibitem{mulcahyetal_2001}
T. M. Mulcahy, J. R. Hull, K. L. Uherka, R. G. Abboud, and J. Juna,
"Test results of 2-kWh flywheel using passive PM and HTS bearings,"
{\it IEEE Trans. Appl. Supercond.}, vol. 11, pp.  1729 - 1734, June 2001.

\bibitem{nagayaetal_2001}
S. Nagaya, N. Kashima, M. Minami, H. Kawashima, and S. Unisuga, "Study
on high- temperature superconducting magnetic bearing for 10 kWh
flywheel energy storage system," {\it IEEE Trans. Appl. Supercond.},
vol. 11, pp. 1649 - 1652, June 2001.

\bibitem{coombsetal_1999} 
T. Coombs, A. M. Campbell, R. Storey, and R. Weller, "Superconducting
magnetic bearings for energy storage flywheels," {\it IEEE
Trans. Appl. Supercond.}, vol. 9, pp. 968 - 971, June 1999.

\bibitem{miyagawaetal_1999}
Y. Miyagawa, H. Kameno, R. Takahata, and H. Ueyama, "A 0.5 kWh
flywheel energy storage system using a high-Tc superconducting
magnetic bearing," {\it IEEE Trans. Appl.  Supercond.}, vol. 9, pp. 996 -
999, June 1999.

\bibitem{dayetal_2002}
A. C. Day, M. Strasik, K. E. McCrary, P. E. Johnson, J. W. Gabrys,
J. R. Schindler, R. A Hawkins, D. L. Carlson, M. D. Higgins, and
J. R. Hull, "Design and testing of the HTS bearing for a 10 kWh
flywheel system," {\it Supercond. Sci. Technol.}, vol. 15, pp. 838 - 841,
2002.

\bibitem{leeetal_1999}
E. Lee, K. B. Ma, T. L. Wilson, and W.-K. Chu, "Characterization of
superconducting bearings for lunar telescopes," {\it IEEE
Trans. Appl. Supercond.}, vol. 9, pp. 911 - 915, June 1999.

\bibitem{hulletal_grav_1999}
J. R. Hull and T. M. Mulcahy, "Gravimeter using high-temperature
superconducting bearing," {\it IEEE Trans. Appl. Supercond.}, vol. 9,
pp. 390 - 393, June 1999.  

\bibitem{chewetal_1995}
A. D. Chew, A. Chambers, and A. P. Toup, "New technique for the
measurement of the intrinsic drag torques of high temperature
superconductor bearings," {\it Appl. Supercond.}, vol.  3, pp. 327 - 338,
1995.


\bibitem{partridge}
R.~B. Partridge, {\em 3K: The Cosmic Microwave Background Radiation}.
\newblock Cambridge University Press, 1995.

\bibitem{mather_etal90}
J. C.~Mather {\em et al.} ``A preliminary measurement of the cosmic microwave 
background spectrum by the Cosmic Background Explorer (COBE) satellite.''
{\it ApJ Lett.} Vol. 354, pp. 37 - 40, 1990

\bibitem{debernardis_etal2000} P.~de Bernardis, {\em et al.} 
``A flat Universe from high-resolution maps of the cosmic microwave 
background radiation'', {\it Nature}, Vol. 404, pp. 955 - 959, 2000. 

\bibitem{hanany_etal2000} S.~Hanany, {\em et al.} 
``MAXIMA-1: a measurement of the cosmic microwave background 
 anisotropy on angular acales of 10' - 5$^{o}$'', {\it ApJ}, Vol. 545, 
pp. L5 - L9, 2000

\bibitem{lee_etal2001} A. T.~Lee, {\em et al.} 
``A high spatial resolution analysis of the MAXIMA-1 
cosmic microwave background anisotropy data'', {\it ApJ}, Vol. 561, 
pp. L1 - L5, 2001.

\bibitem{halverson_etal2002} N.~Halverson, {\em et al.} 
``Degree Angular Scale Interferometer first results: a measurement 
of the cosmic microwave background angular power spectrum'',
{\it ApJ}, Vol. 568, pp. 38 - 45, 2002.

\bibitem{lange_etal2001} A.\ E.~Lange, {\em et al.} 
``Cosmological parameters from the first results of Boomerang'',
{\it Phys. Rev. D} Vol. 63, pp. 042001-1 - 042001-8, 2001

\bibitem{balbi_etal2000}   A.~Balbi, {\em et al.} 
``Constraints on cosmological parameters from MAXIMA-1',
{\it ApJ}, Vol. 545, pp. L1-L5, 2000; Erratum: ibid., Vol. 558, 
pp. L145, 2001

\bibitem{stompor_etal2001} R.~Stompor, {\em et al.} 
``Cosmological implications of the MAXIMA-I high resolution cosmic 
microwave background anisotropy measurement'', {\it ApJ}, 
Vol. 561, pp. L7 - L10, 2001

\bibitem{pryke_etal2002} C.~Pryke, {\em et al.} 
``Cosmological parameter extraction from the first season of
observations with DASI'' {\it ApJ} Vol. 568, pp. 46 - 51, 2002

\bibitem{linde_inflation}
A.\ D.~ Linde, {\it Particle
Physics And Inflationary Cosmology}, Harwood (1990)

\bibitem{Zal97} M.~Zaldarriaga, 
``Polarization of the microwave background in reionized models''
{\it Phys. Rev. D}, Vol. 55, pp. 1822 - 1829, 1997

\bibitem{KamKosSte97a} M.~Kamionkowski, A.~Kosowsky, \&
A.~Stebbins, ``A probe of primordial gravity waves and vorticity''
{\it Phys. Rev. Lett.}, Vol. 78, pp. 2058 - 2061, 1997

\bibitem{lange2002} A.\ E.~Lange,
{\it Proceedings of the Far-IR, Sub-MM, and
MM Detector Technology Workshop}, Monterey, California, April, 2002.

\bibitem{jones_etal88}
Jones, T.J., \& Klebe, D.I., ``A simple infrared polarimeter''
{\it PASP}, Vol. 100, pp. 1158 - 1161, 1988

\bibitem{platt_etal91} Platt, S. R., Hildebrand, R. H., Pernic, J.,
Davidson, J. A., \& Novak, G. ``100-micron array polarimetry from 
the Kuiper Airborne Observatory - instrumentation, techniques, and 
first results'' {\it PASP}, Vol. 103, pp. 1193 - 1210, 1991

\bibitem{leach_etal91} Leach, R. W., Clemens, D. P., Kane, B. D., \&
Barvainis, R., 
``Polarimetric mapping of Orion using MILLIPOL - magnetic activity in BN/KL''
{\it ApJ}, Vol. 370, pp. 257 - 262, 1991

\bibitem{hagmann} C. Hagmann and P.L. Richards, ``Adiabatic
demagnetization refrigerators for small laboratory experiments and
space astronomy,'' {\it Cryogenics}, Vol. 35, No. 5, pp. 303-309, 1995

\bibitem{maxipol}
http://www.physics.umn.edu/maxipol

\bibitem{supercon}
Superconductive Components Inc. 

\bibitem{hulletal_1995} J.\ R.\ Hull, T.\ M.\ Mulcahy,  
K.\ L.\ Uherka and R.\ G.\ Abboud, ``Low rotational 
drag in high-temperature superconducting bearings'', 
{\it IEEE Trans. Appl. Supercon.} Vol. 5, pp. 626 - 629, 1995. 

\bibitem{hulletal_1996} J.\ R.\ Hull, J.\ F.\ 
Labatille and J.\ A.\ Lockwood ``Reduced hysteresis loss in 
superconducting bearings'', {\it Appl. Supercond.} Vol. 4, pp. 1 - 10, 1996. 

\bibitem{zeisberger_gawalek98}
M.\ Zeisberger and W.\ Gawalek, ``Losses in magnetic bearings'',
{\it Mat. Scie. Eng.} Vol. B53, pp. 193-197, 1998.

\bibitem{weinbergeretal_1991}
B.\ R.\ Weinberger, L.\ Lynds, J.\ R.\ Hull, and U.\ Balachandran,
``Low friction in high temperature superconductor bearings'',
{\it Appl. Phys. Lett.} Vol. 59, pp. 1132 - 1134, 1991.

\bibitem{zhangetal_2002}
Y.\ Zhang, Y.\ P.\ Postrekhin, K.\ B.\ Ma, and W.\ K.\ Chu, 
``Reaction wheel with HTS bearings for mini-satellite attitude control'',
{\it Supercond. Sci. Tech.} Vol. 15, pp. 1 - 3, 2002. 



\end{thebibliography}
\end{document}
